\documentclass[12pt]{article}
\usepackage{yfonts}
\usepackage{amsmath}
\usepackage{amssymb}
\usepackage{amstext}
\usepackage{amscd}
\usepackage{amsfonts}

\newcommand{\be}{\begin{equation}}
\newcommand{\nd}{\noindent}
\newcommand{\ee}{\end{equation}}
\newcommand{\ben}{\begin{eqnarray}}
\newcommand{\een}{\end{eqnarray}}

\title{{\bf  Reflections on the q-Fourier transform and the  q-Gaussian function}}

\author{ A. Plastino$^1$ and M. C. Rocca$^{1,\,2}$ \\$^1$
 Instituto de F\'{\i}sica La Plata - CCT-Conicet\\
Universidad Nacional (UNLP) - C.C. 727 (1900) La Plata, Argentina\\
$^2$  Departamento de F\'{\i}sica, Fac. de C. Exactas, UNLP}

\begin{document}

\maketitle

\begin{abstract}

\nd The standard q-Fourier Transform (qFT)  of a
 constant  diverges, which begs for a better
 treatment. In
 addition, Hilhorst has conclusively proved that the ordinary
 qFT is not of a one-to-one character for an infinite set of
 functions [J. Stat. Mech. P10023 (2010)]. Generalizing the ordinary qFT analyzed in [Milan J. Math. {\bf
76} (2008) 307], we appeal here to  a complex q-Fourier transform,
 and show that the problems above mentioned are overcome.

\vskip 3mm

\nd Keywords: q-Fourier transform, tempered ultradistributions,
complex-plane generalization, one-to-one character.

\end{abstract}

\newpage

\renewcommand{\theequation}{\arabic{section}.\arabic{equation}}

\section{Introduction}

\nd    Nonextensive statistical mechanics (NEXT)
\cite{[1],[2],AP}, a well known generalization of the
Boltzmann-Gibbs (BG) one, is used in many scientific and
technological endeavors. NEXT central concept is that of a
nonadditive (though extensive \cite{[3]}) entropic information
measure characterized by the real index q (with q = 1 recovering
the standard BG entropy). Applications include  cold atoms in
dissipative optical lattices \cite{[4]}, dusty plasmas \cite{[5]},
trapped ions \cite{[6]}, spin glasses \cite{[7]}, turbulence in
the heliosphere \cite{[8]}, self-organized criticality \cite{[9]},
high-energy experiments at LHC/CMS/CERN \cite{[10]} and
RHIC/PHENIX/Brookhaven \cite{[11]}, low-dimensional dissipative
maps \cite{[12]}, finance \cite{[13]}, galaxies \cite{AP1},
 and Fokker-Planck equation's studies \cite{AP2}, etc.

\nd    NEXT can be advantageously expressed via q-generalizations
of standard mathematical concepts \cite{borges}. One can mention,
for instance, the logarithm and exponential functions, addition
and multiplication, Fourier transform (FT), the Central Limit Theorem
\cite{tq2}, plane waves, and the representation of the Dirac delta
into plane waves \cite{[15],[16],[17],tq1}.

\nd    Until recently,
 a generic analytical expression for the inverse q-FT for arbitrary
functions and any value of q did not exist \cite{tq4}. This
situation  was adequately remedied in  \cite{PR12}, whose authors,
 by using tempered ultra-distributions \cite{bollini,tp1}, introduced  a complex
q-Fourier transform $F(k,q)$ which exhibits nice properties and is
one-to-one. In turn, this overcame a serious flaw of the original
$F_q-$definition, i.e., not being of the essential one-to-one
nature \cite{tq3}.  Investigations of this kind and related
questions are relevant for field theory and condensed matter
physics, engineering (e.g., image and signal processing), and
mathematical areas for which the standard FT and its inverse play
important roles.

\nd    In this work we focus attention on q-Gaussians, an
essential tool of q-statistics \cite{alf}, that was not discussed
in \cite{PR12}. q-Gaussian behavior is often encountered in quite
distinct settings \cite{alf}. In particular, one has to mention
experimental scenarios in which data are gathered using a set-up
that performs a normalization preprocessing. The ensuing
normalized input, as recorded by the measurement device, will
always be q-Gaussian distributed, if the incoming data exhibit
elliptical symmetry, a rather common feature \cite{alf}. The
q-Fourier transform of the q-Gaussian was discussed in \cite{tq2},
but the corresponding treatment also presented the flaws above
alluded to, a situation  deserving further discussion that we
tackle below.

\setcounter{equation}{0}

\section{Preliminaries}

\nd    It is necessary, before proceeding, to review  materials
developed in \cite{PR12} (more details in the Appendix). So-called
q-exponentials

\be \label{qexp} e_q(x)= [1+(1-q)x]_+^{1/(1-q)},\ee are the
hallmark of Tsallis's statistics \cite{[1]}, being generalizations
of the ordinary exponential functions and  coinciding with them
for $q\rightarrow 1$. Here we will deal with  {\it complex} q-exponentials,
i.e., $e_q(ikx)$ for $1\leq q<2$ with $k$ a real number (see
(\cite{tq1})
\begin{equation}
\label{ep1.1x} e_q(ikx)=[1+i(1-q)kx]^{\frac {1} {1-q}}.
\end{equation}

\nd  Our central tools are {\it distributions}, that is,  linear
functionals that map a set of conventional and well-behaved
functions, called test functions, onto the set of real (complex)
numbers. In this sense, (\ref{ep1.1x}) is to be regarded as a distribution.
{\it Tempered distributions} constitute a subset of the
distributions-set for which the test functions are members of a
special space called Schwartz' one $\mathcal{S}$, a function-space
in which its members possess derivatives that are rapidly
decreasing. $\mathcal{S}$ exhibits a notable  property: {\it the
Fourier transform is an automorphism on}  $\mathcal{S}$, a
property that allows, by duality, to define the Fourier transform
for elements in the dual space of $\mathcal{S}$. This dual is the
space of tempered distributions. In physics it is not uncommon to
face functions that grow exponentially in space or time. In such
circumstances Schwartz' space of tempered distributions is too
restrictive. Instead, ultra-distributions satisfy that need
\cite{bollini}, being continuous linear functionals defined on the
space of entire functions rapidly decreasing on straight lines
parallel to the real axis \cite{bollini}.

\nd An important fact about ultra-distributions is the following:
{\it a tempered distribution is the cut of a tempered
ultra-distribution}. We are not speaking here of  the ``cut" of an
analytic function (see Refs. \cite{tp1,tp2}). Accordingly,
$e_q(ikx)$ is the cut along the real k-axis of a tempered
ultra-distribution \cite{tp1,tp2}, an essential  fact for our
present endeavor,

\begin{equation}
\label{ep1.2x}
E_q(ikx)=\left\{H(x)H[\Im(k)]-H(-x)H[-\Im(k)]\right\}
[1+i(1-q)kx]^{\frac {1} {1-q}},
\end{equation}
with $H(x)$ the Heaviside's step function and $\Im(k)$  the
imaginary part of the complex number $k$.
 The relationship between $e_q(ikx)$ and $E_q(ikx)$
  becomes more clear noting that, if $f(k)$ is a tempered distribution
and $F(k)$ is the corresponding tempered ultra-distribution, then
(\cite{PR12,tp1,tp2})

\begin{equation}
\label{a} f(k)=F(k+i0)-F(k-i0).
\end{equation}
Thus,  (\ref{a}) leads to
\begin{equation}
\label{b} e_q(ikx)=E_q[i(k+i0)x]-E_q[i(k-i0)x]
\end{equation}

\nd At this stage we introduce  the set $\Lambda_{[1,2),\infty}$,
defined as
\begin{equation}
\label{ep1.4x}
{\Lambda}_{[1,2),\infty}=\{f(x)/f(x)\in{\Lambda}_{[1,2),\infty}^+\wedge
f(x)\in{\Lambda}_{[1,2),\infty}^-\},
\end{equation}
where
\[{\Lambda}_{[1,2),\infty}^+=\left\{f(x)/f(x)\{1+i(1-q)kx[f(x)]^{(q-1)}\}^{\frac {1} {1-q}}\in
{\cal L}^1[\mathbb{R}^+]\wedge \right.\]
\begin{equation}
\label{ep1.5x} \left. [f(x)\geq 0;1\leq q<2]\right\}
\end{equation}
and
\[{\Lambda}_{[1,2),\infty}^-=\left\{f(x)/f(x)\{1+i(1-q)kx[f(x)]^{(q-1)}\}^{\frac {1} {1-q}}\in
{\cal L}^1[\mathbb{R}^-]\wedge\right.\]
\begin{equation}
\label{ep1.6x} \left. [f(x)\geq 0;1\leq q<2]\right\}
\end{equation}
With the help of $\Lambda$ and by recourse to    (\ref{ep1.2x})
together with the fact that for a given  $F(k,q)$

\be \label{p2} \lim_{\epsilon \rightarrow
0^+}\,\int\limits_{-\infty}^{\infty}\,dq\,\delta(q-1-\epsilon)\,F(k,q)\,=
F(k) \ee   so that

\be \label{p3} f(x)=
\frac{1}{2\pi}\,\oint\limits_{\Gamma}\,dk\,F(k) e^{-ikx} \ee we
 can define a  complex Umarov-Tsallis-Steinberg (UTS) q-Fourier
transform (of $f(x)\in \Lambda_{[1,2),\infty}$) in the following way
\[F(k,q)=[H(q-1)-H(q-2)]\times \]
\[\left\{H[\Im(k)]\int\limits_0^{\infty} f(x)\{1+i(1-q)kx[f(x)]^{(q-1)}\}^{\frac {1}
{1-q}}, \;dx -\right.\]
\begin{equation}
\label{ep1.3x} \left. H[-\Im(k)]\int\limits_{-\infty}^0 f(x)
\{1+i(1-q)kx[f(x)]^{(q-1)}\}^{\frac {1} {1-q}} \;dx\right\}
\end{equation}
Here $q$ is a real variable such that $1\leq q<2$. {\it The cut
along the real axis of this transform is the real UTS q-Fourier
transform given in \cite{tq2}, \cite{tq1} (see \cite{PR12} for a
simple application of this transform)}. Taking into account that
for $q=1$ the q-Fourier transform is the usual Fourier transform
and using the formula for the inversion of the complex Fourier
transform  straightforwardly leads to the inversion formula for
(\ref{ep1.3x}).

\nd Consider
$$ F(k)=     \lim_{\epsilon \rightarrow 0+}\,\int_1^2\,
\delta(q-1-\epsilon) F(k,q) dq,$$ together with
$$ f(x)= \frac{1}{2\pi}\,\int_{\Gamma}\,dk\,F(k)e^{-ikx}.$$
 Since for $q=1$ our equation (\ref{ep1.3x}) is the complex Fourier
transform
$$  F(k)= H[\Im(k)]\,\int_0^{\infty}\,dx\,f(x)e^{ikx} -
H[-\Im(k)]\,\int_{-\infty}^0\,dx\,f(x)e^{ikx},$$  from
(\ref{ep1.3x}) we find
\begin{equation}
\label{ep1.7x} f(x)=\frac {1}
{2\pi}\oint\limits_{\Gamma}\left[\lim_{\epsilon\rightarrow 0^+}
\int\limits_1^2 F(k,q)\delta (q-1-\epsilon)\;dq\right]
e^{-ikx}\;dk.
\end{equation}
 Eqs.  (\ref{ep1.3x}) and (\ref{ep1.7x}) solve the problem of
inversion of the q-Fourier transform,  which is  of a one-to-one
nature (see \cite{tq3} for fixed $q$). Clearly, from (\ref{a}) and
(\ref{b}), on the real axis, one gets for (\ref{ep1.3x}) and
(\ref{ep1.7x})
\[F(k,q)=[H(q-1)-H(q-2)]\times \]
\begin{equation}
\label{ep1.8x} \int\limits_{-\infty}^{\infty} f(x)
\{1+i(1-q)kx[f(x)]^{(q-1)}\}^{\frac {1} {1-q}} \;dx,
\end{equation}
for the real transform, and
\begin{equation}
\label{ep1.9} f(x)=\frac {1}
{2\pi}\int\limits_{-\infty}^{\infty}\left[\lim_{\epsilon\rightarrow
0^+} \int\limits_1^2 F(k,q)\delta (q-1-\epsilon)\;dq\right]
e^{-ikx}\;dk,
\end{equation}
for its inverse.

\setcounter{equation}{0}

\section{Series expansion of the q-Fourier transform}

\nd Consider now the function

\be
e_q[ikx\,f(x)^{q-1}] \equiv h(x,k,q)=
\{1+(1-q)ikx[f(x)]^{q-1}\}^{\frac {1} {1-q}},
\ee
with $f(x)\in {\Lambda}_{[1,2),\infty}$ , that constitutes
a generalization of  $e_q[ikx]$, whose treatment was discussed
above. Using the series expansions of the logarithm and the
exponential function, we  write
\[e_q[ikx\,f(x)^{q-1}] \equiv  \{1+(1-q)ikx[f(x)]^{q-1}\}^{\frac {1} {1-q}}=
e^{\frac {1} {1-q}\ln\{1+(1-q)ikx[f(x)]^{q-1}\}}=\]
\[e^{\frac {1} {1-q}\sum\limits_{n=1}^{\infty}
\frac {(-1)^{n+1}} {n}(1-q)^n(ikx)^n[f(x)]^{n(q-1)}}=\]
\[e^{\frac {1} {1-q}\sum\limits_{n=1}^{\infty}
\frac {(-1)^{n+1}} {n}(1-q)^n(ikx)^n
e^{n(q-1)\ln f(x)}}=\]
\[e^{\frac {1} {1-q}\sum\limits_{n=1}^{\infty}
\frac {(-1)^{n+1}} {n}(1-q)^n(ikx)^n
\sum\limits_{m=0}^{\infty}\frac {n^m} {m!}
(q-1)^m [\ln f(x)]^m}=\]
\[e^{\left[\sum\limits_{n=1}^{\infty}\sum\limits_{m=0}^{\infty}
\frac {n^{m-1}} {m!} (ikx)^n \ln^m[f(x)](q-1)^{n+m-1}\right]}=\]
Performing the change of variables $n^{'}=m+n$, $m^{'}=m$ and then
making $n^{'}=n$ and $m^{'}=m$ we obtain:
\begin{equation}
\label{ep1.1}
e^{\left\{\sum\limits_{n=0}^{\infty}\left[\sum\limits_{m=0}^{n}
\frac {(n+1-m)^{m-1}} {m!} (ikx)^{n-m+1}
 \ln^m[f(x)]\right](q-1)^n\right\}}.
\end{equation}
Let $g(x,k,n)$ be given by
\begin{equation}
\label{ep1.2} g(x,k,n)=\sum\limits_{m=0}^n \frac {(n-m+1)^{m-1}}
{m!}(ikx)^{n-m+1}\ln^m[f(x)].
\end{equation}
Then,
\begin{equation}
\label{ep1.3}   e_q[ikx\,f(x)^{q-1}] \equiv
\{1+(1-q)ikx[f(x)]^{q-1}\}=h(x,k,q)=e^{\;\sum\limits_{n=0}^{\infty}
g(x,k,n)(q-1)^n}
\end{equation}
or
\begin{equation}
\label{ep1.4} e_q[ikx\,f(x)^{q-1}] \equiv
e^{ikx}e^{\;\sum\limits_{n=1}^{\infty} g(x,k,n)(q-1)^n}.
\end{equation}
According to the  exponential function's expansion  we have
\begin{equation}
\label{ep1.5} e^{\;\sum\limits_{n=1}^{\infty}g(x,k,n)(q-1)^n}=
\sum\limits_{p=0}^{\infty}\frac {\left(\sum\limits_{n=1}^{\infty}
g(x,k,n)(q-1)^n\right)^p} {p!},
\end{equation}
or:
\[e^{\;\sum\limits_{n=1}^{\infty}g(x,k,n)(q-1)^n}=1+
 \sum\limits_{s=1}^{\infty}g(k,x,s) (q-1)^S+\]
\[\frac {1} {2!}
\sum\limits_{s_1=1}^{\infty}\sum\limits_{s_2=1}^{\infty}
g(k,x,s_1)g(k,x,s_2)(q-1)^{s_1+s_2}+\cdot\cdot\cdot+\]
This sum can be rearranged.
Let $l(x,k,n)$ be given by:
\newpage
\[l(x,k,n)=\frac {1} {n!}\sum\limits_{s=n}^{\infty}
\sum\limits_{s_1=1}^{s-n+1}\sum\limits_{s_2=1}^{s-s_1-n+2}
\cdot\cdot\cdot\sum\limits_{s_{n-1}=1}^{s-s_1-s_2-\cdot\cdot\cdot
-s_{n-2}-1}\]
\[g(x,k,s_1)g(x,k,s_2)\ldots
g(x,k,s_{n-1})\cdot\cdot\cdot\]
\begin{equation}
\label{ep1.7}
g(x,k,s-s_1-s_2-\cdot\cdot\cdot-s_{n-1})(q-1)^s
\end{equation}
Then we have:
\begin{equation}
\label{ep1.6} e_q[ikx\,f(x)^{q-1}] \equiv
h(x,k,q)=e^{ikx}\left[1+\sum\limits_{n=1}^{\infty}
l(x,k,n)\right],
\end{equation}
Finally (see Section 2) we can write the q-Fourier transform  of
$e_q[ikx\,f(x)^{q-1}] $ in the fashion
\[F(k,q)=[H(q-1)-H(q-2)]\times \]
\begin{equation}
\label{ep1.8}
\left\{H[\Im(k)]\int\limits_0^{\infty} f(x)h(x,k,q)
\;dx -
H[-\Im(k)]\int\limits_{-\infty}^0 f(x)
h(x,k,q) \;dx\right\}.
\end{equation}

\setcounter{equation}{0}

\section{The q-Fourier transform of the q-Gaussian}

\nd   As stated in the Introduction, our purpose is to calculate
the q-Fourier transform of the q-Gaussian. The calculation is too
involved if one wishes to consider the pertinent expansions up to
arbitrary order on $q-1$. We will content ourselves here with a
first order approach. Accordingly,
\begin{equation}
\label{ep4.1} h(x,k,q)=e^{ikx} [1+g(x,k,1)(q-1)],
\end{equation}
with
\begin{equation}
\label{ep4.2} g(x,k,1)=\frac {(ikx)^2} {2} + ikx \ln[f(x)],
\end{equation}
so that, up  to first order, we have for the q-Fourier transform

\[F(k,q)=[H(q-1)-H(q-2)]\times \]
\[\left\{H[\Im(k)]\int\limits_0^{\infty} \left\{1+
\left\{\frac {(ikx)^2} {2} + ikx \ln[f(x)]\right\}(q-1)\right\}f(x)e^{ikx}
\;dx -\right.\]
\begin{equation}
\label{ep4.3}
\left.H[-\Im(k)]\int\limits_{-\infty}^0
\left\{1+\left\{\frac {(ikx)^2} {2} + ikx \ln[f(x)]\right\}(q-1)\right\}f(x)e^{ikx}
\;dx\right\}
\end{equation}
Let $G(k)$ and $G(k,\beta)$ be given by:
\begin{equation}
\label{ep4.4}
G(k)=\left\{H[\Im(k)]\int\limits_0^{\infty}
f(x)e^{ikx}\;dx -
H[-\Im(k)]\int\limits_{-\infty}^0f(x)e^{ikx}\;dx\right\}
\end{equation}
\begin{equation}
\label{ep4.5} G(k,\beta)=\left\{H[\Im(k)]\int\limits_0^{\infty}
[f(x)]^{\beta}e^{ikx}\;dx -
H[-\Im(k)]\int\limits_{-\infty}^0[f(x)]^{\beta}\begin{large}
\end{large}e^{ikx}\;dx\right\},
\end{equation}
enabling us to write
\[F(k,q)=[H(q-1)-H(q-2)]\times \]
\begin{equation}
\label{ep4.6} G(k)+\left[\frac {k^2} {2}\frac {{\partial}^2}
{\partial k^2}G(k)+k\frac {\partial} {\partial k} \frac {\partial}
{\partial\beta}G(k,\beta)\right]_{\beta=1} (q-1).
\end{equation}
Let $f(x)$ be the q-Gaussian
\begin{equation}
\label{ep4.7} f(x)=C_{q^{'}}[1+ (q^{'}-1)\alpha x^2]^{\frac {1}
{1-q^{'}}},
\end{equation}
where:
\begin{equation}
\label{ep4.8} C_{q^{'}}=\frac {\sqrt{(q^{'}-1)\alpha}}
{B\left(\frac {1} {2}, \frac {1} {q^{'}-1}\frac {1}
{2}\right)}\;\;\;q^{'}\neq 1,
\end{equation}
\begin{equation}
\label{ep4.9} C_1=\sqrt{\frac {\alpha} {\pi}}.
\end{equation}
Thus, using \cite{tt3} we obtain
\[G(k,q^{'})=H[\Im(k)]C_{q^{'}}\frac {\sqrt{\pi}} {2}
\frac {\Gamma\left(\frac {2-q^{'}} {1-q^{'}}\right)}
{[(q^{'}-1)\alpha]^{\frac {1} {1-q^{'}}}}
\left[\frac {2} {(1-q^{'})i\alpha k }\right]^{\frac {2-q^{'}}
{1-q^{'}}-\frac {1} {2}}\times\]
\[\left\{{\bf H}_{\frac {2-q^{'}} {1-q^{'}}-\frac {1} {2}}
\left(\frac {ik} {(1-q^{'})\alpha}\right)-
{\bf N}_{\frac {2-q^{'}} {1-q^{'}}-\frac {1} {2}}
\left(\frac {ik} {(1-q^{'})\alpha}\right)\right\}-\]
\[H[-\Im(k)]C_{q^{'}}\frac {\sqrt{\pi}} {2}
\frac {\Gamma\left(\frac {2-q^{'}} {1-q^{'}}\right)}
{[(q^{'}-1)\alpha]^{\frac {1} {1-q^{'}}}} \left[\frac {2}
{(q^{'}-1)i\alpha k }\right]^{\frac {2-q^{'}} {1-q^{'}}-\frac {1}
{2}}\times\]
\begin{equation}
\label{ep4.10} \left\{{\bf H}_{\frac {2-q^{'}} {1-q^{'}}-\frac {1}
{2}} \left(\frac {ik} {(q^{'}-1)\alpha}\right)- {\bf N}_{\frac
{2-q^{'}} {1-q^{'}}-\frac {1} {2}} \left(\frac {ik}
{(q^{'}-1)\alpha}\right)\right\},
\end{equation}

\[G(k,q^{'},\beta)=H[\Im(k)]C_{q^{'}}^{\beta}\frac {\sqrt{\pi}} {2}
\frac {\Gamma\left(\frac {\beta + 1-q^{'}} {1-q^{'}}\right)}
{[(q^{'}-1)\alpha]^{\frac {\beta} {1-q^{'}}}}
\left[\frac {2} {(1-q^{'})i\alpha k }\right]^{\frac {\beta +1-q^{'}}
{1-q^{'}}-\frac {1} {2}}\times\]
\[\left\{{\bf H}_{\frac {\beta+1-q^{'}} {1-q^{'}}-\frac {1} {2}}
\left(\frac {ik} {(1-q^{'})\alpha}\right)-
{\bf N}_{\frac {\beta+1-q^{'}} {1-q^{'}}-\frac {1} {2}}
\left(\frac {ik} {(1-q^{'})\alpha}\right)\right\}-\]
\[H[-\Im(k)]C_{q^{'}}^{\beta}\frac {\sqrt{\pi}} {2}
\frac {\Gamma\left(\frac {\beta +1-q^{'}} {1-q^{'}}\right)}
{[(q^{'}-1)\alpha]^{\frac {\beta} {1-q^{'}}}} \left[\frac {2}
{(q^{'}-1)i\alpha k }\right]^{\frac {\beta +1-q^{'}}
{1-q^{'}}-\frac {1} {2}}\times\]
\begin{equation}
\label{ep4.11} \left\{{\bf H}_{\frac {\beta+1-q^{'}}
{1-q^{'}}-\frac {1} {2}} \left(\frac {ik}
{(q^{'}-1)\alpha}\right)- {\bf N}_{\frac {\beta+1-q^{'}}
{1-q^{'}}-\frac {1} {2}} \left(\frac {ik}
{(q^{'}-1)\alpha}\right)\right\},
\end{equation}
where ${\bf H}$ and ${\bf N}$ are the Struve and Neumann
functions, respectively. The q-Fourier transform of the q-Gaussian
is now
\[F(k,q,q^{'})=[H(q-1)-H(q-2)]\times \]
\begin{equation}
\label{ep4.12} G(k,q^{'})+\left[\frac {k^2} {2}\frac
{{\partial}^2} {\partial k^2}G(k,q^{'})+k\frac {\partial}
{\partial k} \frac {\partial}
{\partial\beta}G(k,q^{'},\beta)\right]_{\beta=1} (q-1).
\end{equation}
The cuts on the real axis of $G(k,q^{'})$ and $G(k,q^{'},\beta)$
are
\[G(k,q^{'})=C_{q^{'}}\frac {\sqrt{\pi}} {2}
\frac {\Gamma\left(\frac {2-q^{'}} {1-q^{'}}\right)}
{[(q^{'}-1)\alpha]^{\frac {1} {1-q^{'}}}}
\left[\frac {2} {(1-q^{'})i\alpha (k+i0) }\right]^{\frac {2-q^{'}}
{1-q^{'}}-\frac {1} {2}}\times\]
\[\left\{{\bf H}_{\frac {2-q^{'}} {1-q^{'}}-\frac {1} {2}}
\left(\frac {i(k+i0)} {(1-q^{'})\alpha}\right)-
{\bf N}_{\frac {2-q^{'}} {1-q^{'}}-\frac {1} {2}}
\left(\frac {i(k+i0)} {(1-q^{'})\alpha}\right)\right\}+\]
\[C_{q^{'}}\frac {\sqrt{\pi}} {2}
\frac {\Gamma\left(\frac {2-q^{'}} {1-q^{'}}\right)}
{[(q^{'}-1)\alpha]^{\frac {1} {1-q^{'}}}}
\left[\frac {2} {(q^{'}-1)i\alpha (k-i0) }\right]^{\frac {2-q^{'}}
{1-q^{'}}-\frac {1} {2}}\times\]
\begin{equation}
\label{ep4.13} \left\{{\bf H}_{\frac {2-q^{'}} {1-q^{'}}-\frac {1}
{2}} \left(\frac {i(k-i0)} {(q^{'}-1)\alpha}\right)- {\bf
N}_{\frac {2-q^{'}} {1-q^{'}}-\frac {1} {2}} \left(\frac {i(k-i9)}
{(q^{'}-1)\alpha}\right)\right\},
\end{equation}
\[G(k,q^{'},\beta)=C_{q^{'}}^{\beta}\frac {\sqrt{\pi}} {2}
\frac {\Gamma\left(\frac {\beta + 1-q^{'}} {1-q^{'}}\right)}
{[(q^{'}-1)\alpha]^{\frac {\beta} {1-q^{'}}}}
\left[\frac {2} {(1-q^{'})i\alpha (k+i0) }\right]^{\frac {\beta +1-q^{'}}
{1-q^{'}}-\frac {1} {2}}\times\]
\[\left\{{\bf H}_{\frac {\beta+1-q^{'}} {1-q^{'}}-\frac {1} {2}}
\left(\frac {i(k+i0)} {(1-q^{'})\alpha}\right)-
{\bf N}_{\frac {\beta+1-q^{'}} {1-q^{'}}-\frac {1} {2}}
\left(\frac {i(k+i0)} {(1-q^{'})\alpha}\right)\right\}+\]
\[C_{q^{'}}^{\beta}\frac {\sqrt{\pi}} {2}
\frac {\Gamma\left(\frac {\beta +1-q^{'}} {1-q^{'}}\right)}
{[(q^{'}-1)\alpha]^{\frac {\beta} {1-q^{'}}}} \left[\frac {2}
{(q^{'}-1)i\alpha (k-i0) }\right]^{\frac {\beta +1-q^{'}}
{1-q^{'}}-\frac {1} {2}}\times\]
\begin{equation}
\label{ep4.14} \left\{{\bf H}_{\frac {\beta+1-q^{'}}
{1-q^{'}}-\frac {1} {2}} \left(\frac {i(k-i0)}
{(q^{'}-1)\alpha}\right)- {\bf N}_{\frac {\beta+1-q^{'}}
{1-q^{'}}-\frac {1} {2}} \left(\frac {i(k-i0)}
{(q^{'}-1)\alpha}\right)\right\}.
\end{equation}
Now,  the real q-Fourier transform takes the form
\[F(k,q,q^{'})=[H(q-1)-H(q-2)]\times \]
\begin{equation}
\label{ep4.15} G(k,q^{'})+\left[\frac {k^2} {2}\frac
{{\partial}^2} {\partial k^2}G(k,q^{'})+k\frac {\partial}
{\partial k} \frac {\partial}
{\partial\beta}G(k,q^{'},\beta)\right]_{\beta=1} (q-1).
\end{equation}

\setcounter{equation}{0}

\section{The q-Fourier transform of the q-Gaussian for  fixed q}

In this section we will provide an alternative path to the
computation of the q-Fourier transform,  for a q-Gaussian, in the
case of a  fixed q-value. Note that previously, an interesting
calculation of this transformation, for the real axis, has been
presented  in Ref.  \cite{tr1}. \vskip 3mm \nd We start with
\[F(k,q)=H[\Im(k)]\int\limits_0^{\infty} C_q[1+(q-1)\alpha x^2]^{
\frac {1} {1-q}}\times\]
\[\left\{1+(1-q)ikx\left\{C_q[1+(q-1)\alpha x^2]^{\frac {1} {1-q}}
\right\}({q-1}\right\}^{\frac {1} {1-q}}\;dx-\]
\[H[-\Im(k)]\int\limits_{-\infty}^0 C_q[1+(q-1)\alpha x^2]^{
\frac {1} {1-q}}\times\]
\begin{equation}
\label{ep5.1} \left\{1+(1-q)ikx\left\{C_q[1+(q-1)\alpha
x^2]^{\frac {1} {1-q}} \right\}(q-1)\right\}^{\frac {1}
{1-q}}\;dx,
\end{equation}
$1\leq q <2$. Simplifying terms we obtain
\[F(k,q)=H[\Im(k)]\int\limits_0^{\infty} C_q
[(q-1)\alpha x^2+e^{-\frac {i\pi} {2}}(q-1)C_q^{q-1}kx+1 ]^{\frac
{1} {1-q}}\;dx-\]
\begin{equation}
\label{ep5.2} H[-\Im(k)]\int\limits_{-\infty}^0 C_q [(q-1)\alpha
x^2+e^{\frac {i\pi} {2}}(q-1)C_q^{q-1}kx+1 ]^{\frac {1}
{1-q}}\;dx.
\end{equation}
Effecting the change of variables $\sqrt{(q-1)\alpha}\;x=y$ the
q-Fourier transform adopts the appearance
\[F(k,q)=\frac {H[\Im(k)]} {\sqrt{(q-1)\alpha}}
\int\limits_0^{\infty} C_q
\left[y^2+e^{-\frac {i\pi} {2}}C_q^{q-1}\sqrt{\frac {q-1} {\alpha}}ky+1
\right]^{\frac {1} {1-q}}\;dy-\]
\begin{equation}
\label{ep5.3} \frac {H[-\Im(k)]} {\sqrt{(q-1)\alpha}}
\int\limits_0^{\infty} C_q \left[y^2+e^{\frac {i\pi}
{2}}C_q^{q-1}\sqrt{\frac {q-1} {\alpha}}ky+1 \right]^{\frac {1}
{1-q}}\;dy.
\end{equation}
Using now the result given in \cite{tt5} (${\bf P}_\nu^\mu$ is the
associated Legendre function)
\[{\bf P}_\nu^\mu(z)=\frac {2^\mu \Gamma(1-2\mu)
(z^2-1)^{\frac {\mu} {2}}} {\Gamma(1-\mu)\Gamma(-\mu-\nu)
\Gamma(\nu-\mu+1)}\times\]
\begin{equation}
\label{ep5.4}
\int\limits_0^\infty\left(1+2tz+t^2\right)^{\mu-\frac {1} {2}}
t^{-1-\nu-\mu}\;dt,
\end{equation}
and we can write:
\begin{equation}
\label{ep5.5}
\int\limits_0^\infty\left(1+2tz+t^2\right)^{\mu-\frac {1} {2}}
\;dt=\Gamma(-\mu) 2^{-\mu-1} (z^2-1)^{\frac {\mu}
{2}}{\bf P}_{-\mu-1}^\mu(z),
\end{equation}
where:
\begin{equation}
\label{ep5.6} \gamma=\frac {C_q^{q-1}} {2} \sqrt{\frac {q-1}
{\alpha}} \;\;\;\; \mu=\frac {1} {1-q}+\frac {1} {2},
\end{equation}
so that
\[F(k,q)=C_q\frac {\Gamma(-\mu)} {\sqrt{(q-1)\alpha}}
2^{-\mu-1}e^{-\frac {i\pi\mu} {2}} (\gamma^2 k^2 + 1)^{\frac {
\mu} {2}}\times\]
\begin{equation}
\label{ep5.7} \left\{H[\Im(k)]{\bf P}_{-1-\mu}^{\mu}(e^{-\frac
{i\pi} {2}}\gamma k) -H[-\Im(k)]e^{i\pi\mu} {\bf
P}_{-1-\mu}^{\mu}(e^{\frac {i\pi} {2}}\gamma k)\right\},
\end{equation}
which is the q-Fourier transform of the q-Gaussian on the complex
plane for  fixed q.

\nd   Our next step is to evaluate the q-Fourier transform of the
q-Gaussian on the real axis.  This needs calculating the cut of
(\ref{ep5.7}) along the real axis. Using \cite{tt6} we reach the
equality
\[(\gamma^2 k^2 + 1)^{\frac {\mu} {2}}
\left\{H[\Im(k)]{\bf P}_{-1-\mu}^{\mu}(e^{-\frac {i\pi} {2}}\gamma k)
-H[-\Im(k)]e^{i\pi\mu}
{\bf P}_{-1-\mu}^{\mu}(e^{\frac {i\pi} {2}}\gamma k)\right\}=\]
\[H[\Im(k)]\frac {(\gamma k + i)^{\mu}} {\Gamma(1-\mu)}
F\left(-\mu,1+\mu,;1-\mu;\frac {1+i\gamma k} {2}\right)-\]
\begin{equation}
\label{ep5.8} H[-\Im(k)]e^{i\pi\mu} \frac {(\gamma k - i)^{\mu}}
{\Gamma(1-\mu)} F\left(-\mu,1+\mu,;1-\mu;\frac {1-i\gamma k}
{2}\right)=C(k,q),
\end{equation}
where $F$ is the hypergeometric function.

\nd   The cut $c(k,q)$ along the real axis of $C(k,q)$ is
\[c(k,q)=\frac {(\gamma k + i)^{\mu}} {\Gamma(1-\mu)}
F\left(-\mu,1+\mu,;1-\mu;\frac {1+i\gamma k} {2}\right)+\]
\begin{equation}
\label{ep5.9} e^{i\pi\mu} \frac {(\gamma k - i)^{\mu}}
{\Gamma(1-\mu)} F\left(-\mu,1+\mu,;1-\mu;\frac {1-i\gamma k}
{2}\right).
\end{equation}
Now, according to \cite{tt7}
\begin{equation}
\label{ep5.10} F(-\mu, 1+\mu; 1-\mu; z)=(1-z)^{-\mu}F(1, -2\mu;
1-\mu; z),
\end{equation}
and then
\[c(k,q)=\frac {2^{\mu}e^{\frac {i\pi\mu} {2}}} {\Gamma(1-\mu)}
\left[F\left(1, -2\mu; 1-\mu,\; \frac {1+i\gamma k} {2}\right)\right.+ \]
\begin{equation}
\label{ep5.11} \left.F\left(1, -2\mu; 1-\mu,\; \frac {1-i\gamma k}
{2}\right)\right],
\end{equation}
so that via  the result \cite{tt8} we have
\[F\left(1, -2\mu; 1-\mu; \frac {1+i\gamma k} {2}\right)=
-F\left(1, -2\mu; 1-\mu; \frac {1-i\gamma k} {2}\right)+\]
\begin{equation}
\label{ep5.12} \frac {2\Gamma(1-\mu) \sqrt{\pi}} {\Gamma(\frac {1}
{2}-\mu)} (1+{\gamma}^2 k^2)^{\mu}.
\end{equation}
 Eqs. (\ref{ep5.11}) and (\ref{ep5.12}) lead to  the cut
$f(k,q)$ of $F(k,q)$ in (\ref{ep5.7})
\begin{equation}
\label{ep5.13} f(k,q)=\left(1 + \frac {C_q^{2(q-1)}(q-1)k^2}
{4\alpha}\right)^{ \frac {1} {1-q} + \frac {1} {2}},
\end{equation}
which is the q-Fourier transform of the q-Gaussian on the real
axis for  fixed q. From (\ref{ep5.13}) we see that the q-Fourier
transform of a q-Gaussian is another q'-Gaussian with
\[q^{'}= 1 - \frac {2(1-q)} {3+q}\]
This result can also be obtained  using the
real q-Fourier transform
\[f(k,q)=\int\limits_{-\infty}^{\infty} C_q
[1+(q-1)\alpha x^2]^{\frac {1} {1-q}}\times \]
\begin{equation}
\label{ep5.14} \left\{1+(1-q)ikx\left\{C_q[1+(q-1)\alpha
x^2]^\frac {1} {1-q}\right\}^{q-1}\right\}^\frac {1} {1-q}\;dx,
\end{equation}
$1\leq q <2$.

\section*{Conclusions}

\nd By recourse to tempered ultra-distributions we have evaluated
a complex q-Fourier transform $F(k,q)$ for the q-Gaussian that is
one-to-one, solving thus a flaw of the original $F_q-$definition,
i.e., not being of the essential one-to-one nature, as
illustrated, for instance, in \cite{PR12}. The computation has
been done both for a floating $q$ and for a fixed one. An
essential piece of the Tsallis' machinery has thus been rebuilt.

\section{Appendix: Tempered Ultradistributions and
Distributions of Exponential Type }

\setcounter{equation}{0}

For the benefit of the reader we give a brief summary of the main
properties of distributions of exponential type and tempered
ultra-distributions.

\nd {\bf Notations}. The notations are almost textually taken from
Ref. \cite{tp2}. Let $\boldsymbol{{\mathbb{R}}^n}$ (res.
$\boldsymbol{{\mathbb{C}}^n}$) be the real (resp. complex)
n-dimensional space whose points are denoted by
$x=(x_1,x_2,...,x_n)$ (resp $z=(z_1,z_2,...,z_n)$). We shall use
the notations:

(a) $x+y=(x_1+y_1,x_2+y_2,...,x_n+y_n)$\; ; \;
    $\alpha x=(\alpha x_1,\alpha x_2,...,\alpha x_n)$

(b)$x\geqq 0$ means $x_1\geqq 0, x_2\geqq 0,...,x_n\geqq 0$

(c)$x\cdot y=\sum\limits_{j=1}^n x_j y_j$

(d)$\mid x\mid =\sum\limits_{j=1}^n \mid x_j\mid$

\nd Let $\boldsymbol{{\mathbb{N}}^n}$ be the set of n-tuples of
natural numbers. If $p\in\boldsymbol{{\mathbb{N}}^n}$, then
$p=(p_1, p_2,...,p_n)$, and $p_j$ is a natural number, $1\leqq
j\leqq n$. $p+q$ stands for $(p_1+q_1, p_2+q_2,..., p_n+q_n)$ and
$p\geqq q$ means $p_1\geqq q_1, p_2\geqq q_2,...,p_n\geqq q_n$.
$x^p$ entails $x_1^{p_1}x_2^{p_2}... x_n^{p_n}$. We shall denote
by $\mid p\mid=\sum\limits_{j=1}^n p_j $ and call $D^p$  the
differential operator
${\partial}^{p_1+p_2+...+p_n}/\partial{x_1}^{p_1}
\partial{x_2}^{p_2}...\partial{x_n}^{p_n}$

\nd For any natural $k$ we define $x^k=x_1^k x_2^k...x_n^k$ and
${\partial}^k/\partial x^k= {\partial}^{nk}/\partial x_1^k\partial
x_2^k...\partial x_n^k$

\nd The space $\boldsymbol{{\cal H}}$  of test functions such that
$e^{p|x|}|D^q\phi(x)|$ is bounded for any $p$ and $q$, being
defined [see Ref. (\cite{tp2})] by means of the countably set of
norms
\begin{equation}
\label{ep2.1} {\|\hat{\phi}\|}_p=\sup_{0\leq q\leq p,\,x} e^{p|x|}
\left|D^q \hat{\phi} (x)\right|\;\;\;,\;\;\;p=0,1,2,...
\end{equation}

\nd The space of continuous linear functionals defined on
$\boldsymbol{{\cal H}}$ is the space
$\boldsymbol{{\Lambda}_{\infty}}$ of the distributions of the
exponential type given by ( ref.\cite{tp2} ).
\begin{equation}
\label{ep2.2} T=\frac {{\partial}^k} {\partial x^k} \left[
e^{k|x|}f(x)\right]
\end{equation}
where $k$ is an integer such that $k\geqq 0$ and $f(x)$ is a
bounded continuous function. \nd In addition we have
$\boldsymbol{{\cal H}}\subset\boldsymbol{{\cal S}}
\subset\boldsymbol{{\cal S}^{'}}\subset
\boldsymbol{{\Lambda}_{\infty}}$, where $\boldsymbol{{\cal S}}$ is
the Schwartz space of rapidly decreasing test functions
(ref\cite{tp6}).

\nd   The Fourier transform of a function $\hat{\phi}\in
\boldsymbol{{\cal H}}$ is
\begin{equation}
\label{ep3.1} \phi(z)=\frac {1} {2\pi}
\int\limits_{-\infty}^{\infty}\overline{\hat{\phi}}(x)\;e^{iz\cdot
x}\;dx
\end{equation}
According to ref.\cite{tp2}, $\phi(z)$ is entire analytic and
rapidly decreasing on straight lines parallel to the real axis. We
shall call $\boldsymbol{{\cal H}}$ the set of all such functions.
\begin{equation}
\label{ep3.2} \boldsymbol{{\cal H}}={\cal
F}\left\{\boldsymbol{{\cal H}}\right\}
\end{equation}
The topology in $\boldsymbol{{\cal H}}$ is defined by the
countable set of semi-norms:
\begin{equation}
\label{ep3.4} {\|\phi\|}_{k} = \sup_{z\in V_k} |z|^k|\phi (z)|,
\end{equation}
where $V_k=\{z=(z_1,z_2,...,z_n)\in\boldsymbol{{\mathbb{C}}^n}:
\mid Im z_j\mid\leqq k, 1\leqq j \leqq n\}$

\nd The dual of $\boldsymbol{{\cal H}}$ is the space
$\boldsymbol{{\cal U}}$ of tempered ultra-distributions [see Ref.
(\cite{tp2} )]. In other words, a tempered ultra-distribution is a
continuous linear functional defined on the space
$\boldsymbol{{\cal H}}$ of entire functions rapidly decreasing on
straight lines parallel to the real axis. \nd Moreover, we have
$\boldsymbol{{\cal H}}\subset\boldsymbol{{\cal S}}
\subset\boldsymbol{{\cal S}^{'}}\subset \boldsymbol{{\cal U}}$.

\nd   $\boldsymbol{{\cal U}}$ can also be characterized in the
following way [see Ref. (\cite{tp2} )]: let $\boldsymbol{{\cal
A}_{\omega}}$ be the space of all functions $F(z)$ such that:

\nd   ${{\boldsymbol{A)}}}$- $F(z)$ is analytic for $\{z\in
\boldsymbol{{\mathbb{C}}^n} : |Im(z_1)|>p,
|Im(z_2)|>p,...,|Im(z_n)|>p\}$.

\nd   ${{\boldsymbol{B)}}}$- $F(z)/z^p$ is bounded continuous  in
$\{z\in \boldsymbol{{\mathbb{C}}^n} :|Im(z_1)|\geqq
p,|Im(z_2)|\geqq p, ...,|Im(z_n)|\geqq p\}$, where $p=0,1,2,...$
depends on $F(z)$.

\nd Let $\boldsymbol{\Pi}$ be the set of all $z$-dependent
pseudo-polynomials, $z\in \boldsymbol{{\mathbb{C}}^n}$. Then
$\boldsymbol{{\cal U}}$ is the quotient space

\nd   ${{\boldsymbol{C)}}}$- $\boldsymbol{{\cal
U}}=\boldsymbol{{\cal A}_{\omega}/\Pi}$

\nd By a pseudo-polynomial we understand a function of $z$ of the
form $\;\;$ $\sum_s z_j^s G(z_1,...,z_{j-1},z_{j+1},...,z_n)$ with
$G(z_1,...,z_{j-1},z_{j+1},...,z_n)\in\boldsymbol{{\cal
A}_{\omega}}$

\nd Due to these properties it is possible to represent any
ultra-distribution as [see Ref. (\cite{tp2} )]
\begin{equation}
\label{ep3.6} F(\phi)=<F(z), \phi(z)>=\oint\limits_{\Gamma} F(z)
\phi(z)\;dz
\end{equation}
$\Gamma={\Gamma}_1\cup{\Gamma}_2\cup ...{\Gamma}_n,$  where the
path ${\Gamma}_j$ runs parallel to the real axis from $-\infty$ to
$\infty$ for $Im(z_j)>\zeta$, $\zeta>p$ and back from $\infty$ to
$-\infty$ for $Im(z_j)<-\zeta$, $-\zeta<-p$. ($\Gamma$ surrounds
all the singularities of $F(z)$).

\nd Eq. (\ref{ep3.6}) will be our fundamental representation for a
tempered ultra-distribution. Use is also  made of the ``Dirac
formula" for ultra-distributions [see Ref. (\cite{tp1})]
\begin{equation}
\label{ep3.7} F(z)=\frac {1} {(2\pi
i)^n}\int\limits_{-\infty}^{\infty} \frac {f(t)}
{(t_1-z_1)(t_2-z_2)...(t_n-z_n)}\;dt
\end{equation}
where the ``density'' $f(t)$ is such that
\begin{equation}
\label{ep3.8} \oint\limits_{\Gamma} F(z) \phi(z)\;dz =
\int\limits_{-\infty}^{\infty} f(t) \phi(t)\;dt.
\end{equation}
While $F(z)$ is analytic on $\Gamma$, the density $f(t)$ is in
general singular, so that the r.h.s. of (\ref{ep3.8}) should be
interpreted in the sense of distribution theory.

\nd Another important property of the analytic representation is
the fact that on $\Gamma$, $F(z)$ is bounded by a power of $z$
\cite{tp2}
\begin{equation}
\label{ep3.9} |F(z)|\leq C|z|^p,
\end{equation}
where $C$ and $p$ depend on $F$.

\nd The representation (\ref{ep3.6}) implies that the addition of
a pseudo-polynomial $P(z)$ to $F(z)$ does not alter the
ultra-distribution:
\[\oint\limits_{\Gamma}\{F(z)+P(z)\}\phi(z)\;dz=
\oint\limits_{\Gamma} F(z)\phi(z)\;dz+\oint\limits_{\Gamma}
P(z)\phi(z)\;dz\] However,
\[\oint\limits_{\Gamma} P(z)\phi(z)\;dz=0.\]
As $P(z)\phi(z)$ is entire analytic in some of the variables $z_j$
(and rapidly decreasing), we obtain:
\begin{equation}
\label{ep3.10} \oint\limits_{\Gamma} \{F(z)+P(z)\}\phi(z)\;dz=
\oint\limits_{\Gamma} F(z)\phi(z)\;dz.
\end{equation}

\vspace{0.5 in}

\nd {\bf Acknowledgments}
The authors thank Prof. C. Tsallis for
having called their attention to the present problem.

\end{document}